\documentstyle[12pt]{article}
\begin{document}

\newcommand\ie {{\it i.e. }}
\newcommand\eg {{\it e.g. }}
\newcommand\etc{{\it etc. }}
\newcommand\cf {{\it cf.  }}
\newcommand\viz{{\it viz. }}
\newcommand\grad{\nabla}
\newcommand\noi{\noindent}
\newcommand\seq{\;\;=3D\;\;}
\newcommand\barcaps{\cal}
\newcommand\jump{\vspace*{17pt}}
\newcommand\emptypage{~~~ \eject}
\setlength{\baselineskip}{17pt}
\def\be{\begin{eqnarray}}
\def\ee{\end{eqnarray}}
\newenvironment{draftequation}[1]{\be\label{#1}}{\ee}
\newcommand\bbe[1]{\begin{draftequation}{#1}}
\newcommand\eee{\end{draftequation}}
\def\ps{p\hspace{-0.075in}/}
\def\pis{\pi\hspace{-0.075in}/}
\def\half{{\textstyle{1 \over 2}}}
\def\ihalf{{\textstyle{i \over 2}}}
\def\D{{\cal D}}
\newcommand\art[1]{\cite{#1}}
\newcommand\ekv[1]{(\ref{#1})}

\def\aligneq#1#2{\cr\cr\noalign{\hbox to \hsize{#1\hfil}\vskip-3\baselineskip
    \hbox to \hsize{\hfil(\arabic{equation})}\addtocounter{equation}{1}
    \vskip2\baselineskip\llap{\hbox to .5cm{\hfil}}}\label{#2}}

\begin{flushright}
USITP-94-08\\
October  1994
\end{flushright}
\bigskip
\Large
\begin{center}
\bf{Hamiltonian BRST Quantization of the Conformal String}\\

\bigskip

\normalsize
by\\
\bigskip

H.Gustafsson, U.Lindstr\"om, P.Saltsidis, B.Sundborg and R.v.Unge\\
{\it ITP\\
University of Stockholm\\
Box 6730, Vanadisv\"agen 9\\
S-113 85 Stockholm\\
SWEDEN}\\
\end{center}
\vspace{1.0cm}
\normalsize

{\bf Abstract:} We present a new formulation of the tensionless string ($T=
0$) where the space-time conformal symmetry is manifest. Using a Hamiltonian
BRST scheme we quantize this {\em Conformal String} and find that it has
critical dimension $D=2$. This is in keeping with our classical result that
the model describes massless particles in this dimension. It is also
consistent with our previous results which indicate that quantized
conformally symmetric tensionless strings describe a topological phase away
{}from $D=2$.

We reach our result by demanding nilpotency of the
BRST charge and consistency with the Jacobi identities. The derivation
is presented in two different ways: in operator language and using
mode expansions.

Careful attention is payed to regularization, a crucial ingredient in our
calculations.

\eject

\begin{flushleft}
\large
{\bf Introduction}
\end{flushleft}

The high-energy limit of strings has been studied with regard to
scattering \art{AMAT3}-\art{GROS4} as well to the high-temperature
behaviour \art{ALVA1}-\art{ATIC1}, but it is far from fully
understood. Open problems are the understanding of the high-energy
symmetries of Gross \art{GROS1} and Moore \art{Moore}, and the relation to the
conjectured ``topological phase'' of general covariance
\art{WITT1},\art{WITT2}.

The zero tension limit ($T\to 0$) of strings and
superstrings, \art{jiulbs}-\art{sc} provide a possible high energy limit
of the corresponding tensile ($T\ne 0$) models\footnote{See, e.g.,
\art{ulbsgt1} \art{BigT}.}.
They are also interesting in their own right, since they provide new, albeit
somewhat degenerate, string models. In addition their quantization is
sufficiently different from the ($T\ne 0$) models to provide new insights into
the quantization of extended objects.

In a previous article, \art{BigT}, the condition under which the space-time
conformal symmetry of the bosonic tensionless string survives quantization
was investigated. The surprising conclusion is that this symmetry holds good
at the quantum level essentially only if the physical states of the theory
are space-time diffeomorphism singlets, indicating that the theory describes
a topological string phase.

The treatment in \art{BigT} is based on an action with only the Poincar\'e
subgroup of the space-time conformal symmetry manifest. Furthermore the
quantization is carried out in a light-cone gauge describing only physical
degrees of freedom. Thus, all of the space-time conformal symmetry has to be
explicitly checked. In view of the surprising outcome it is important to
corroborate the results in \art{BigT} using other methods. A first step in
this direction is to identify the obstructions to quantization using
different quantization schemes.

In this paper we present a Hamiltonian BRST quantization of the $T\to 0$
limit of the bosonic string, starting from {\em the Conformal String}, (named
in analogy to the conformal particle in \art{RMBN}), a $D+2$ dimensional
formulation which is classically equivalent to that used in \art{BigT}, but
where the space-time conformal symmetry {\em is manifest}\footnote{In $D=2$
only invariance under the finite dimensional m\"obius subgroup of the
conformal group is manifest.}. We find that this theory has critical
dimension $D=2$, a result which is consistent with \art{BigT} where the $D-2$
transversal degrees of freedom are the basic objects and where hence $D\ne 2$
{}from the outset. It is also in keeping with the classical results presented
in the present paper that our model describes massless particles in $D=2$. As
was shown in \art{BigT}, the conformal invariance for massless particles in
$D>2$ survives quantization and one would expect that to be true in $D=2$
also\footnote{This is independent of the usual difficulties with masslessness
in $D=2$.}. Finally, the space-time conformal group is infinite-dimensional
in $D=2$ which is also expected to give good quantum behaviour.

\bigskip
The content of the paper is as follows: \\

\smallskip
In Section $1$ we present the classical Conformal String
theory, (the $D+2$ dimensional action with manifest conformal symmetry), its
symmetries, equations of motion and some of their consequences. In particular
we discuss the relation to other formulations,
the classical picture as a set of conformal particles \art{RMBN} obeying a
constraint, the Hamiltonian description in terms of the classical constraints
and their algebra and the classical BRST charge.\\

\smallskip
Section $2$ is the main part of our paper and contains the quantum
theory. It starts out with a discusssion of the vacuum (2.1-2). In
many ways the $T=0$ string behaves like a collection of particles, and
the  vacuum we
find appropriate is indeed annihilated by the momentum
operators. Starting from this requirement we find the full vacuum
which accomodates the existence of ghosts and should allow for finite inner
products in analogy to
\art{MarnBRST}.
The key problem in our calculations is to keep track of possible
divergencies. The method for doing this is to introduce a regularized
delta function in the canonical commutation-relations and then to
choose a particular ``physical'' ordering of the coordinate and momentum
operators in the calculations involving the composite expressions for the
constraints. This prescription and its application in investigating the
nilpotency of the quantum BRST charge is contained in (2.3-4), concluding
with the
discovery of the critical dimension $D=2$.

In (2.5) we set out anew with slightly different approach. Here we use
a mode expansion of the operators and constraints and regulate
infinite sums rather than delta functions. We
discuss the central extensions of the quantum constraint algebra,
introduce ``extended constraints'', i.e., include the ghost sector in
the constraints, and derive the consequences of the central extensions
for the nilpotency of the BRST charge in an economic way. Here the
effect on the Jacobi identities is also discussed. We discover that
the Jacobi identities hold if and only if $D=2$ and that the BRST charge is
then nilpotent.\\

\smallskip
In Section 3 we reexamine the classical theory in the critical
dimension $D=2$, discuss our results, mentioning how we believe that the
present results can be reconciled with the results of \art{BigT} away from
$D=2$, and point to some future topics for
investigation.\\

\smallskip
In Appendices A and B we have collected some explicit calculations
along with a presentation of how the critical dimension $D=26$ for
$T\ne 0$ strings is
derived using our methods.

\begin{flushleft}
\section{The classical theory}
\end{flushleft}
The ``Conformal String Theory'' we consider is given by the action
\bbe{COFA}
S=\int{d^2\xi \left(V^\alpha V^\beta \D _\alpha X^M \D _\beta X^N
\eta_{MN} +\Phi X^2\right)}
\eee
where $X^M(\xi ),\quad M=0,...,D+1$ is an embedding of the world
sheet, coordinatized by $(\xi^\alpha ) = (\xi ^0,\xi ^1) \equiv (\tau
,\sigma)$, into the target space with metric
\bbe{METR}
\eta _{MN} = \left( \matrix{\eta_{mn}\quad  \hfill 0 \quad 0  \cr
 0...0 \quad \hfill 1 \quad 0 \cr 0...0 \quad \hfill 0 -1 \cr}
\right).
\eee

Here $\eta_{mn}, \quad m=0,...,D-1,$ is the metric in $D$-dimensional
Minkowski space, which shows that we have written the theory in a
$D+2$ dimensional space with signature $(-+++......+-)$. Furthermore
$V^\alpha$ is a contravariant vector-density
(whose transformation properties will be given below) and
the scale-covariant derivatives $\D_\alpha$ are given by
\bbe{SDER}
\D_\alpha \equiv \partial_\alpha + W_\alpha
\eee
with $W_\alpha$ being the gauge field for scale transformations.
Finally $X^2 \equiv X^MX^N\eta_{MN}$ and $\Phi$ is a scalar density
Lagrange multiplier field that restricts the theory to the $D+2$
dimensional light cone. The model is reminiscent of the conformal particle
\art{RMBN}, hence its name.

The action (\ref{COFA}) is a $D+2$ dimensional version of the action
first used in \cite{ulbsgt1} and subsequently employed in
investigations of the $T\to 0$ limit of strings \cite{ulbsgt2, liro,
jiulbs, BigT}. Just like the slightly different action in \cite{akul}
the space-time conformal symmetry has been made manifest by adding one
time-like and one space-like dimension.  In Hamiltonian form this
theory was also treated in \cite{ISBE}.

The symmetries of the action (\ref{COFA}) are two dimensional (world sheet)
diffeomorphisms, local $D+2$ scale transformations\footnote{Not
to be confused with the $D$-dimensional dilatations of the conformal
group, which are included in the ($D+2$)-dimensional rotations.}, an
``additional'' local
two dimensional
symmetry and global ($D+2$)-dimensional rotations. Explicitly, they are given
by:\\
\begin{flushleft}(i) Diffeomorphisms ($\epsilon = \epsilon (\xi)$):
\end{flushleft}
\bbe{DIFF}
\delta_\epsilon X^M&=&\epsilon ^\alpha \partial_\alpha  X^M \equiv
\epsilon \cdot \partial X^M\cr
\delta_\epsilon V^\alpha &=& \epsilon \cdot \partial  V^\alpha
-V\cdot\partial\epsilon^\alpha +
\half(\partial \cdot\epsilon )V^\alpha\cr
\delta_\epsilon W_\alpha &=&\epsilon \cdot \partial W_\alpha
+W_\beta \partial_\alpha\epsilon^\beta\cr
\delta_\epsilon \Phi &=& \partial_\alpha (\epsilon^\alpha \Phi)
\aligneq{(ii) Scale transformations ($\lambda = \lambda (\xi)$):}{SCAL}
\delta_\lambda X^M &=& \lambda X^M \cr
\delta_\lambda V^\alpha &=& -\lambda V^\alpha\cr
\delta_\lambda W_\alpha &=& -\partial _\alpha \lambda\cr
\delta_\lambda \Phi &=& -2\lambda \Phi
\aligneq{(iii) Additional symmetry ($\Xi_\alpha = \Xi_\alpha (\xi))$:}{ADD}
\delta_\Xi X^M &=& 0\cr
\delta_\Xi V^\alpha &=&0\cr
\delta_\Xi W_\alpha &=& -\Xi_\alpha\cr
\delta_\Xi \Phi &=& 2(V\cdot W)(V\cdot \Xi )-\partial_\alpha (V^\alpha
V\cdot \Xi )
\aligneq{(iv) Rotations ($\Lambda_{MN} \equiv \Lambda_{[MN]}$):}{ROT}
\delta_\Lambda X^M &=& \Lambda _N ^M X^N \cr
\delta_\Lambda V^\alpha &=& 0\cr
\delta_\Lambda W_\alpha &=& 0\cr
\delta_\Lambda \Phi &=& 0
\eee

\bigskip

The field equations that result from the action (\ref{COFA}) are:
\bbe{FEQN}
\delta V^\alpha :\quad &&\D_\alpha X^M V^\beta \D_\beta X_M=0\cr
\delta X^M :\quad &&\D_\alpha (V^\alpha V^\beta \D_\beta X_M) - \Phi X_M=0\cr
\delta W_\alpha : \quad && V^\alpha V^\beta X^M \D_\beta X_M =0\cr
\delta \Phi : \quad && X^2 =0,
\eee
where, in the second line,
$\D_\alpha (\ )^\alpha =(\partial_\alpha - W_\alpha ) (\ )^\alpha$,
the change in sign being due to the scaling property of the term it
acts on.

In a reparametrization gauge $V^\alpha =(E^{-\half}(\tau),0)$ the
equations (\ref{FEQN}) become
\bbe{PART}
X^2 &=& 0\cr
\dot{X}^2 &=&0\cr
\ddot{X}^M &=&{\dot{E}\over E}\dot{X}^M+\left({\tilde{\Phi}E^2+W^2+
{\dot{E}\over E}W-\dot{W}}\right) X^M\cr
\dot{X}^MX'_M &=& 0
\eee
where $\tilde{\Phi}\equiv \Phi E^{-1}$ and $W\equiv W_0$. Here dot
denotes $\tau$- and prime denotes $\sigma$- derivatives.
For fixed
$\sigma$, the three first equations in (\ref{PART}) are precisely the equations
of motion for the conformal particle with action \cite{RMBN}:

\bbe{PACT}
S=\int{d\tau \left({E^{-1}(\dot{X}+W)^2+E\Phi X^2}\right)}.
\eee
Hence the conformal string may be viewed (in this gauge) as a
collection of conformal particles, one at each $\sigma$, subject to a
constraint, (the last equation in (\ref{PART})). Also in the Minkowski
space formulation of the zero tension limit of the bosonic, spinning
and superstrings there are similar gauge choices where massless
particle, spinning particle and superparticle equations may be recognized.

Integrating out $W_\alpha$ and fixing a scaling gauge, the action
(\ref{COFA}) can be reduced to the space-time action employed in,
e.g., \cite{BigT},
\bbe{LDAC}
S=\int{d^2\xi}V^\alpha \partial _\alpha X^mV^\beta \partial_\beta
X_m,
\eee
where the Minkowski metric $\eta _{mn}$ is used in the $X$ summation.

In the remaining part of this letter we will be interested in the
Hamiltonian form of the theory. It is given by the Hamiltonian
\bbe{HAM}
H=\lambda_i\phi^i, \quad i=-1,0,1,L
\eee
where $\lambda_i$ are Lagrange multiplier fields{\footnote{The
Lagrange multipliers correspond to the fields $V$, $W$ and $\Phi$ in
the Lagrange formulation.}} and the constraints
$\phi^i$ are given by:
\bbe{CON}
\phi^{-1}&=&P^2\cr
\phi^0&=&P_MX^M\cr
\phi^1&=&X^2\cr
\phi^L&=&P_M{X'}^M,
\eee
with $X^M$ and $P_N$ fulfilling the usual canonical
relations
\bbe{comereal}
\left\{{X^M(\sigma ),P_N(\sigma ' )}\right\}&=&\delta^M_N\delta
  (\sigma -\sigma ').
\eee
These constraints are all first class and form the following algebra,
\bbe{CALG}
\left\{{\phi^1(\sigma ),\phi^{-1}(\sigma ' )}\right\}&=&2\left({\phi^0(\sigma )
+\phi^{0}(\sigma ' )}\right)\delta(\sigma - \sigma ')\cr
\left\{{\phi^1(\sigma ),\phi^{0}(\sigma ' )}\right\}&=&\left({\phi^1(\sigma )
+\phi^{1}(\sigma ' )}\right)\delta(\sigma - \sigma ')\cr
\left\{{\phi^0(\sigma ),\phi^{-1}(\sigma ' )}\right\}&=&\left({\phi^{-1}
    (\sigma )
+\phi^{-1}(\sigma ' )}\right)\delta(\sigma - \sigma ')\cr
\left\{{\phi^L(\sigma ),\phi^{L}(\sigma ' )}\right\}&=&\left({\phi^L(\sigma )
+\phi^{L}(\xi ' )}\right)\delta '(\xi - \xi ')\cr
\left\{{\phi^1(\sigma ),\phi^{L}(\sigma ' )}\right\}&=&\left({-\phi^1(\sigma )
+\phi^{1}(\sigma ' )}\right)\delta '(\sigma - \sigma ')\cr
\left\{{\phi^0(\sigma ),\phi^{L}(\sigma ' )}\right\}&=&\phi^0(\sigma )
  \delta '(\sigma - \sigma ')\cr
  \left\{{\phi^{-1}(\sigma ),\phi^{L}(\sigma ' )}\right\}&=&\left({\phi^{-1}
      (\sigma )
+\phi^{-1}(\sigma ' )}\right)\delta '(\sigma - \sigma ').
\eee
All other Poisson brackets are zero. Here $\phi^{L}$ generates a
Virasoro algebra and $\phi ^i$ transform under this algebra with
conformal spin $1-i$. The whole algebra is a semi direct product
between the Virasoro algebra and a $SU(1,1)$ Ka\v c-Moody algebra, both
without central extensions. Note that the subalgebra formed by $\phi^L$ and
$\phi^{-1}$ is isomorphic to the gauge algebra in the Minkowski formulation
of the tensionless string \art{BigT}. With the structure constants of
the constraint algebra at hand we may write down the classical
Hamiltonian BRST charge $\cal Q$ \cite{FRAD,ROBP}:
\bbe{Q}
{\cal{Q}} = \int{d\sigma} \left({\phi ^ic^i + 4b^0c^{-1}c^1+2b^1c^0c^1
+2b^{-1}c^{-1}c^0+
\partial b^1c^Lc^1}\right.\cr
+\left.{b^{-1}\partial c^{-1}c^L+b^0\partial c^0c^L+b^L\partial
c^Lc^L}\right).
\eee
Here we have introduced the (anti-) ghosts $(b^i)$, $c^i$, corresponding to
the constraints $\phi^i$, fulfilling the canonical relations
\be
\left\{ b^{i}(\sigma),c^{j}(\sigma^{\prime})\right\}^{+}=-i\delta^{ij}
  \delta(\sigma-\sigma^{\prime}).
\ee
The couplings are determined by the structure constants $f^{ijk}$ of the
algebra\ekv{CALG} according to the general prescription of \art{FRAD}:
\bbe{eqnhere}
 {\cal Q}=\phi^i c^i - \half f^i_{jk}b_ic^jc^k.
\eee
The classical nilpotency, ${\cal Q}^2=0$, is guaranteed by
construction.  Whether this survives in the quantum theory is the
topic of the rest of this article.

\bigskip
\section{The quantum theory}

\subsection{The Vacuum}
We define the matter part of the vacuum $|0\rangle_{p}$ by the condition that
\bbe{Pvac}
 P^{M}(\xi)|0\rangle_{p}=0\;\;\forall M .
\eee
In terms of their Fourier components this reads
\be
p^{M}_{n}|0\rangle_{p}=0\;\;\forall M,n,
\ee
which because of the commutation relations implies
\be
x^{M}_{n}|0\rangle_{p}\neq 0 \;\; \forall M,n .
\ee
We have arrived at these definitions by a wish to keep a relation to the
$T\neq 0$ Hilbert space and vacuum. We have argued as follows:

{}From the expressions of the oscillators of the closed tensile string, with
$T$ denoting the tension,
\bbe{OscXP}
 \alpha^{M}_{n}(T) = -in\sqrt{T}x^{M}_{n}+\frac{1}
{2\sqrt{T}}p^{M}_{n}\nonumber\\ \tilde{\alpha}^{M}_{-n}(T) =
in\sqrt{T}x^{M}_{n}+ \frac{1}{2\sqrt{T}}p^{M}_{n},\nonumber
\eee
and the requirement on the tensile vacuum
\bbe{RecVac}
  \alpha^{M}_{n}(T)|0\rangle_{T}=\tilde{\alpha}^{M}_{n}
(T)|0\rangle_{T}=0\hspace{1cm}\forall n > 0,\nonumber
\eee
we find that
\bbe{Tvac}
\left(-i|n|\sqrt{T}x^{M}_{n}+\frac{1}{2\sqrt{T}}p^{M}_{n}\right)
|0\rangle_{T}=0\hspace{1cm}\forall n\neq 0.
\eee
Should we assume a similar equation to hold also for a $T$ independent
tensionless vacuum we would be forced to choose $ x^{M}_{n}|0\rangle
=p^{M}_{n}|0\rangle =0$ for all $M$ and $n\neq 0$. This is inconsistent with
the commutation relations. A possible modification of this is that only the
positive modes annihilate the vacuum. However, this corresponds to the $T\to
0$ limit of a tensile theory with left and right oscillators treated
differently. As we have no reason to suspect such a breaking of symmetry, we
now turn to the one remaining choice, advocated in \cite{BigT}. In
(\ref{Tvac}) we see that the $P$ operators become more and more important as
$T\rightarrow 0$. We thus choose a vacuum, $|0\rangle_{p}$ which is
annihilated by $P$ operators only. Then
\be
\left(-i|n|\sqrt{T}x^{M}_{n}+\frac{1}{2\sqrt{T}}p^{M}_{n}\right)
  |0\rangle_{p}\rightarrow 0\hspace{1cm}\forall n\neq 0,
  \ee
when $T\rightarrow 0$ and the $|0\rangle_{p}$ vacuum satisfies the
$T\rightarrow 0$ limit of the vacuum conditions of the tensile theory in a way
consistent with the canonical commutation relations. An additional
complication compared to \cite{BigT} is that there are two extra dimensions.
One might entertain the idea that operators acting on these dimensions should
be treated differently; however, this would mean breaking of the manifest
space-time conformal covariance and, since we want to examine if this symmetry
is preserved in the quantized theory, this is not a convenient choice. Our
final choice of the matter part of the vacuum is thus
(\ref{Pvac}).

\subsection{The full vacuum}
Since the full theory involves ghosts we will also have to choose vacuum
states for these. Our guiding principle in search of a viable ghost vacuum
is that the total ghost and matter vacuum state should be physical, and thus
be annihilated by the BRST charge.

To make our manipulations well defined, we have to work in a space
with finite inner products. In \cite{MarnBRST} it was shown that this can be
achieved by introducing an additional state space together with a well defined
bilinear form. Using this formalism, we will have {\em bra} and {\em ke}t
states belonging to {\em different} state spaces such that $\langle bra\;
sector | ket\; sector \rangle = finite$
\footnote{In \cite{InnerProd}, a proposal for a general BRST invariant inner
  product of physical states using this formalism is given. However, one
  should be aware that in \cite{InnerProd} only systems with a finite number
  of degrees of freedom are treated. We believe that the case of infinite
  number of degrees of freedom can be dealt with in the same way using the
  regularization methods introduced in this article.}.

Following the prescription of \cite{MarnBRST}, we will take
the {\em ket } states to be built from our vacuum of choice, $|0\rangle_{p}$,
and the {\em bra } states to be built from $\mbox{}_{x}\langle 0|$ satisfying
$\mbox{}_{x}\langle 0|0\rangle_{p}=1$. Since the theory contains ghosts,
we will also have to consider vacuum states for these, so that the {\em ket}
ghost vacuum is given by $|G\rangle$ and the {\em bra} ghost vacuum is given
by $\langle G^{\prime}|$, satisfying $\langle G^{\prime}|G\rangle =1$.
{}From our choice of vacuum, and from the requirement that the BRST
charge (\ref{Q}) should annihilate the vacuum, we find that also
$\{{\cal Q},P^{M}\}=2iX^{M}c^{1}+iP^{M}c^{0}-i\partial (P^{M}c^{L})$
must annihilate the vacuum. Commutation relations tells us that
$X^{M}$ cannot annihilate the vacuum. Therefore one has to impose
$c^{1}|G\rangle=0$. This means that $\langle G^{\prime}|b^{1}=0$.  Similarly
we find that $\langle G^{\prime} |c^{-1}=0$ and $b^{-1}|G\rangle =0$.

To summarize we have
states built from the following vacuum:
\bbe{FullVac}
 |0\rangle=|0\rangle_{p}|G\rangle\nonumber\\ \langle 0|=\mbox{}_{x}\langle 0|
  \langle G^{\prime}|,\nonumber
\eee
which satisfies
\bbe{FullVacCond}
  P^{M}|0\rangle =c^{1}|0\rangle =b^{-1}|0\rangle=0\nonumber\\ \langle
0|X^{M}=\langle 0|c^{-1}=\langle 0|b^{1}=0\nonumber\\
\langle0|0\rangle = \mbox{}_{x}\langle 0|0\rangle_{p}\langle
G^{\prime}|G\rangle=1.\nonumber
\eee
The action of $c^{L},b^{L},c^{0},b^{0}$ on the vacuum is, for now,
left undetermined.
\smallskip

\subsection{Regularization}
To keep track of possible divergencies in our calculations we have to
regulate. This is done as in \cite{BigT}, using an approximate delta function
which fulfills
\bbe{RegD}
 \lim_{\epsilon\rightarrow 0}\int d\sigma
f(\sigma)\delta_{\epsilon}(\sigma) =f(0) &
\delta_{\epsilon}(-\sigma)=\delta_{\epsilon}(\sigma)\nonumber\\ \int
d\sigma\delta_{\epsilon}(\sigma)=1 &
\delta_{s\epsilon}(\sigma)=\frac{1}{s}\delta_{\epsilon}(\frac{\sigma}{s}).
\eee
In the limit $\epsilon\rightarrow 0$, the regularized delta
function approaches a real Dirac delta function. The regularisation is
achieved by introducing the approximate delta function in the
canonical commutation relations,
\bbe{RegComm}
\left[X^{M}(\sigma),P^{N}(\sigma^{\prime})\right] & = &
  i\eta^{MN}\delta_{\epsilon}(\sigma-\sigma^{\prime})\nonumber\\
  \left\{\{b(\sigma),c(\sigma^{\prime})\right\} & = &
    \delta_{\epsilon}(\sigma-\sigma^{\prime}),
\eee
which amounts to adding terms which vanish as $\epsilon\rightarrow 0$ to the
commutator. This formalism will allow us to isolate infinities appearing in our
calculations, they will come out as terms diverging as we let $\epsilon$ go
to zero.

We want to investigate if there are anomalies in the quantized version of the
constraints (\ref{CALG}) and also if there are quantum obstructions to the
nilpotency of the quantum BRST charge.

To this end we have to calculate commutators of composite operators and use
that the fundamental fields satisfy (\ref{RegComm}). The result of these
computations is in general not a local object but it may be reinterpreted as an
$\epsilon$ expansion in local quantities.

In doing this we expect to uncover possible infinities, i.e., terms
proportional to $\frac{1}{\epsilon}$. It is thus crucial to control the
$\epsilon$ dependence in these calculations. Before giving our prescription for
obtaining this control, we illustrate the situation by way of two examples.

Consider first the distributional equivalence
\bbe{SimpleCorr}
A(\sigma)\delta_{\epsilon}(\sigma-\sigma^{\prime})&=&A(\sigma^{\prime})
\delta_{\epsilon}(\sigma-\sigma^{\prime})\\&&+\frac{b\epsilon^{2}}{2}
(A^{\prime\prime}(\sigma^{\prime})\delta_{\epsilon}(\sigma-\sigma^{\prime})
-2A^{\prime}(\sigma^{\prime})\delta_{\epsilon}^{\prime}(\sigma-
\sigma^{\prime}))+{\cal O}(\epsilon^{4}),\nonumber
\eee
where $b$ is a constant
\bbe{bdef}
 b=\int d\sigma\delta_{1}(\sigma)\sigma^{2}.
\eee
To verify this equivalence one has to use test functions and integrate with
respect to $\sigma$ and $\sigma^{\prime}$, and then use the scaling properties
of the regularized delta function to bring out all $\epsilon$ corrections
explicitly. Note that for the relation (\ref{SimpleCorr}) to be a true
$\epsilon$ expansion, we have to require that $A$ and the derivatives of $A$
are well behaved when $\epsilon\rightarrow 0$ such that there
are no hidden divergencies from these fields.

Our constraints (\ref{CON}) are composite operators upon quantization. As
usual this leads to ordering ambiguities. This has two consequences: The
regularized expressions for the constraints may contain $\cal O(\epsilon)$
terms and a reordering of the fundamental fields in the constraints may
generate $\frac{1}{\epsilon}$ terms. In the language of the above example
(\ref{SimpleCorr}) the $A$'s and derivatives of $A$'s are {\em not} well
behaved. This may lead to problems, as is spelled out in more detail
in our second example:

A relation similar
to (\ref{SimpleCorr}), but involving composite operators, is
\bbe{PSplit}
  2X(\sigma)\cdot
P(\sigma^{\prime})\delta_{\epsilon}(\sigma-\sigma^{\prime})
=\left\{{X(\sigma)\cdot P(\sigma)+X(\sigma^{\prime})\cdot
P(\sigma^{\prime})}\right\}\delta_{\epsilon}(\sigma-\sigma^{\prime})
\nonumber\\
-\frac{b\epsilon^{2}}{2}\left\{{X^{\prime}(\sigma)\cdot P^{\prime}(\sigma)
  +X^{\prime}(\sigma^{\prime})\cdot
P^{\prime}(\sigma^{\prime})}\right\}\delta_
{\epsilon}(\sigma-\sigma^{\prime})\\
+\frac{b\epsilon^{2}}{2}\left\{{X(\sigma)\cdot P^{\prime}(\sigma)+X(\sigma^
{\prime})\cdot
P^{\prime}(\sigma^{\prime})}\right.
\nonumber\\
\left.{ -X^{\prime}(\sigma)\cdot P(\sigma)-X^{\prime}(\sigma^{\prime})\cdot
    P(\sigma^{\prime}) }\right\}
\delta_{\epsilon}^{\prime}(\sigma-\sigma^{\prime})+ {\cal
O}(\epsilon^{4})\nonumber.
\eee

The extra terms appear to vanish when $\epsilon$ goes to zero and we are
left with the usual relation. However, if we reorder
all terms, from $XP$ ordering to $PX$ ordering we get
$\frac{1}{\epsilon}$ contributions from {\em all} terms since the
commutator of $X$ and $P$ evaluated at the same point is
\bbe{deltazero}
\left[X^{M}(\sigma),P^{N}(\sigma)\right]=i\eta^{MN}\delta_{\epsilon}(0)
=i\eta^{MN}\frac{1}{\epsilon}\delta_{1}(0),
\eee
and each derivative of $X$ or $P$ will bring out an extra
$\pm\frac{1}{\epsilon}$.\\

The two examples above reveal the problems we are faced with in trying to
control the $\epsilon$ dependence: A reordering may give non-trivial
corrections and we have to be very careful in choosing the ordering, not to
have a hidden $\epsilon$ dependence. As illustrated in the first example,
we may avoid such a dependence if all functionals of $X$ and $P$ are bounded
in the limit $\epsilon\rightarrow 0$. We define such a bounded operator, with
all $X$ operators and their derivatives appearing to the left of all $P$
operators and their derivatives, to be {\em physically ordered}. Then the
space of all states with smooth momentum dependence can be handled by studying
$\epsilon$-dependence as above. From the definition of the full vacuum we also
find that physical ordering will mean that all $b^{1}$ operators appear to the
left of all $c^{1}$ operators and that all $c^{-1}$ operators appear to the
left of all $b^{-1}$ operators. The ordering of the $c^{0},b^{0},c^{L},b^{L}$
ghosts is not determined since we have not determined the ghost vacuum states
corresponding to these fields. This is very similar what one does for the
ordinary string. There one orders all operators with positive modes
to the right of negative modes, to make sure that they annihilate the
vacuum. In our case however, we are forced to choose a different
vacuum, which in turn forces us to the present construction.\\

Our scheme for keeping track of the $\epsilon$ dependence
is thus as follows; We start from the ordinary hermitean
expressions and then use physical ordering in our
calculations. We perform the calculations with a nonzero $\epsilon$
which allows us to take care of possible ordering constants in a
consistent way. At the end, we let the regularization parameter go to zero.

It turns out that only for $D=2$ is the quantum theory thus obtained well
defined.

\smallskip

\subsection{The BRST anomaly}

In this subsection we calculate ${\cal Q}^2$ in the quantum theory using the
operator quantization procedure described above.

In the quantum theory ${\cal Q}|phys\rangle=0$ and $\langle phys|{\cal
Q}=0$. In particular, these equations hold true for the vacuum states. To
make these equations well defined we have to physically order the BRST charge.
To make sure that the BRST charge is hermitean we start from ${\cal Q}_{H}
=\half({\cal Q}+{\cal Q}^{\dagger})$, putting all terms in physical order,
using the regularization introduced before. The full hermitean ${\cal
Q}$, including the $\frac{1}{\epsilon}$ corrections from reordering, reads
\bbe{qBRST}
 {\cal Q}_{P}&=&\int d\sigma \left(
P^{2}c^{-1}+X\cdot Pc^{0}+X^{2}c^{1}+
X^{\prime}\cdot Pc^{L}-4ic^{-1}b^{0}c^{1}+2ib^{1}c^{0}c^{1}\right.\nonumber\\
&&+2ic^{-1}c^{0}b^{-1}+i\partial b^{1}c^{L}c^{1}+ i\partial
c^{-1}c^{L}b^{-1}+ib^{0}\partial c^{0}c^{L}\\ &&\left. +ib^{L}\partial
c^{L}c^{L}+\frac{(2-D)ia}{2\epsilon}c^{0}+\frac{ia}
{2\epsilon}\partial c^{L}+{\cal O}(\epsilon) \right),\nonumber\\
a&\equiv &\delta_{1}(0),\nonumber
\eee
where the choice of ordering for the last two cubic ghost terms is
left undetermined. We observe that the last ordering correction term is
a surface term and we may therefore subsequently ignore it.
\smallskip

We now examine the nilpotency of the physically ordered BRST charge
${\cal Q}_{P}$ using  $2{\cal Q}_{P}^{2}=\left\{ {\cal
Q}_{P}, {\cal Q}_{P}\right\}$, and keeping track of all possible $\epsilon$
corrections. We find
\bbe{QSqrCalc}
 2{\cal Q}_{P}^{2}=\frac{(2-D)a}{\epsilon}\int
d\sigma\left(4c^{1}c^{-1}+ c^{0}\partial
c^{L}\right)+\frac{(D-2)A}{\epsilon}\int d\sigma 4c^{1}c^{-1},
\eee
where
\bbe{Adef}
  A=\int d\mu \delta_{1}(\mu)\delta_{1}(\mu).
\eee
This vanishes for $D=2$, indicating the possibility of a consistent
two dimensional quantum theory. In the course of the calculation we also found
that the ordering of the $c^{0},b^{0},c^{L},b^{L}$ terms do not affect this
result.

\subsection{Central extensions and consistency with the Jacobi identities}

In this subsection  we investigate the consequencies of the
Jacobi identities for the constraint algebra and reexamine the nilpotency
of the BRST-charge using a mode
expansion of the operators.

We shall consider closed strings. Letting $F$ denote any of the
coordinates $P^{M}(\sigma),X^{M}(\sigma)$ we define the Fourier modes
$f^{n}_{m}$ by the decomposition
\be
F^{M}(\sigma)=\frac{1}{\sqrt{\pi}} \sum_{-\infty}^{+\infty}f_{n}^{M}e^{-2in
  \sigma}
\ee
so that the non-vanishing Poisson brackets for the coordinate modes
are
\bbe{xp1}
\{x^{M}_{m},p^{N}_{n}\}_{P.B.}=\eta^{MN}\delta_{m+n}.
\eee
The Fourier modes of the constraints read
\be
\phi^{-1}_{m}&=&\half \sum_{-\infty}^{+\infty}p_{k}\cdot p_{m-k}=
0\label{ccon1}\\
\phi^{0}_{m}&=&\half \sum_{-\infty}^{+\infty}x_{k}\cdot p_{m-k}=
0\label{ccon2}\\
\phi^{1}_{m}&=&\half \sum_{-\infty}^{+\infty}x_{k}\cdot x_{m-k}=
0\label{ccon3}\\
\phi^{L}_{m}&=&-i\sum_{-\infty}^{+\infty}kx_{k}\cdot p_{m-k}=
0\label{ccon4}
\ee
Notice that we have multiplied all the constraints by the constant
$\frac{\sqrt{\pi}}{2}$ to make the notation simpler. They satisfy the algebra
\be
\{\phi^{a}_{m},\phi^{L}_{n}\}_{P.B.}&=&-i(m+a n)\phi^{a}_{m+n}\label{cacon1}\\
\{\phi^{L}_{m},\phi^{L}_{n}\}_{P.B.}&=&-i(m-n)\phi^{L}_{m+n}\label{cacon2}\\
\{\phi^{a}_{m},\phi^{b}_{n}\}_{P.B.}&=&(1-\delta_{ab})(a-b)
\phi^{a +b}_{m+n}\label{cacon3}
\ee
and all the other brackets vanish.

The mode ex\-pansions of the ghosts as\-soci\-ated with the con\-straints
$\phi^{a}(\sigma)$ and $\phi^{L}(\sigma)$ satisfy the following fundamental
Poisson bracket relation
\bbe{bc}
\{c^{i}_{m},b^{j}_{n}\}^{+}_{P.B.}=-i\delta_{m+n}\delta^{ij}.
\eee

Using
the relations (\ref{cacon1})-(\ref{cacon3}) we obtain the
expression for the BRST charge

\bbe{Q2}
{\cal Q}_C&=& \sum_{k}
(\phi_{-k}^{1}c_{k}^{1}+\phi_{-k}^{0}c_{k}^{0}+\phi_{-k}^{-1}c_{k}^{-1}+
\phi^{L}_{-k}c_{k}^{L})\nonumber\\
&+&
\sum_{k,l}-[2ic_{-k}^{1}c_{-l}^{-1}b_{k+l}^{0}+ic_{-k}^{1}c_{-l}^{0}
b_{k+l}^{1}-ic_{-k}^{-1}c_{-l}^{0}b_{k+l}^{-1}+\nonumber\\
& & (k+l)c_{-k}^{1}c_{-l}^{L}b_{k+l}^{1}+(k-l)c_{-k}^{-1}c_{-l}^{L}
b_{k+l}^{-1}+kc_{-k}^{0}c_{-l}^{L}b_{k+l}^{0}+\nonumber\\
& & \half (k-l)c_{-k}^{L}c_{-l}^{L}b_{k+l}^{L}].
\eee
It can be checked that this {\it classical} charge has the desired
property
\be
\{{\cal Q}_C,{\cal Q}_C\}=0.
\ee

To quantize the system we have to replace Poisson brackets
by commutators according to
$i\{\hspace{.1in}\}_{(P.B.)\pm}\to[\hspace{.1in}]_{\pm}\hspace{.08in}
(\hbar\equiv 1)$.  Then (\ref{xp1}) and (\ref{bc}) become
\bbe{com}
[x^{M}_{m},p^{N}_{n}]=i\delta_{m+n} \eta^{MN},\qquad
[b^{i}_{m},c^{j}_{n}]=\delta^{ij}_{m+n}
\eee

Since, in the quantum theory $x^{M}_{m},p^{M}_{m},c^{i}_{m}$ and
$b^{i}_{m}$ are non-commuting operators, one must resolve ordering ambiguities
in the constraints that contain products of these operators. Since
$x^{M}_{k}$ commutes with $p^{M}_{m-k}$ unless $m=0$, we see from
(\ref{ccon1})-(\ref{ccon4}) that such ambiguities arise only in the
expressions for $\phi^{0}_{0}$ and $\phi^{L}_{0}$.

As we have no natural way
of resolving these ambiguities yet, we simply
define $\hat{\phi}^{0}_{0}$ and $\hat{\phi}^{L}_{0}$ to be given by some
definite ordered expressions $ \hat{\phi}^{0}_{0}\equiv\!
\;\; :\phi^{0}_{0}\! :$ and $\hat{\phi}^{L}_{0}\equiv\! \;\;
:\phi^{L}_{0}\!:$, where the ``hat'' $\hat{}$  denotes an abstract operator.
In the classical theory the constraints must vanish for the allowed motions
of the string.  Hence in the quantum theory we demand that a physical state
$\left| phys\right\rangle $ satisfy the following conditions
\be
(\hat{\phi}^{0}_{0}-\alpha_{0}) \left| phys\right\rangle &\equiv &
(:\!\phi^{0}_{0}\! :-\alpha_{0})\left| phys\right\rangle=0\label{ord1}\\
(\hat{\phi}^{L}_{0}-\alpha_{L}) \left| phys\right\rangle &\equiv &
(:\!\phi^{L}_{0}\!:-\alpha_{L})\left| phys\right\rangle =0\label{ord2},
\ee
where because of ordering ambiguities we include an ordering constant.
For the definite {\it physical ordering} these constants take the values
discussed in Appendix A.

Let us now look at the constraint algebra. The right hand sides of
equation (\ref{cacon1}), for $a=0=m+n$, and equation (\ref{cacon3}),
for $a+b=0=m+n$, can be expressed in terms of $\phi^{0}_{0}$. But when
expressing the right hand side in terms of a definite $\hat{\phi}^{0}_{0}$ we
have to take ordering corrections into account. The same is true for the
$\hat{\phi}^{L}_{0}$ operator.  So, in the quantum case, the constraint
algebra takes the form
\be
\left[ \hat{\phi}^{1}_{m},\hat{\phi}^{-1}_{n}\right] &=&
  2i\hat{\phi}^{0}_{m+n}+d^{1,-1}_{m}\delta_{m+n}\cr
  \left[ \hat{\phi}^{L}_{m},\hat{\phi}^{L}_{n}\right] &=&
    (m-n)\hat{\phi}^{L}_{m+n}+d^{L,L}_{m}\delta_{m+n}\cr
    \left[ \hat{\phi}^{0}_{m},\hat{\phi}^{L}_{n}\right] &=&
      m\hat{\phi}^{0}_{m+n}+d^{0,L}_{m}\delta_{m+n}\cr
      \left[ \hat{\phi}^{0}_{m},\hat{\phi}^{0}_{n}\right] &=&
        d^{0,0}_{m}\delta_{m+n}\cr
        \left[ \hat{\phi}^{1}_{m},\hat{\phi}^{0}_{n}\right] &=&
          i\hat{\phi}^{1}_{m+n}\cr
          \left[ \hat{\phi}^{-1}_{m},\hat{\phi}^{0}_{n}\right] &=&
            -i\hat{\phi}^{-1}_{m+n}\cr
            \left[ \hat{\phi}^{1}_{m},\hat{\phi}^{L}_{n}\right] &=&
              (m+n)\hat{\phi}^{1}_{m+n}\cr
              \left[ \hat{\phi}^{-1}_{m},\hat{\phi}^{L}_{n}\right] &=&
                (m-n)\hat{\phi}^{-1}_{m+n}
\ee
where all the constraints are assumed to be ordered. In these
relations we have included a possible central extension for the
commutator $\left[\hat{\phi}^{0}_{m},\hat{\phi}^{0}_{n}\right]$ since
this one contains the "dangerous" operator $\hat{\phi}^{0}_{m}$ twice.

The quantum version of the classical algebra (\ref{cacon1})-(\ref{cacon3}),
thus contains central extensions and can be written in a
general notation
\be\label{gcomm}
\left[\hat{\Psi}^{a},\hat{\Psi}^{b}\right]=U^{ab}_{c}\hat{\Psi}^{c}+d^{ab}.
\ee
The values of additional structure constants $d^{ab}$ are constrained
by the Jacobi identities
\be
\left[\left[\hat{\Psi}^{a},\hat{\Psi}^{b}\right],\hat{\Psi}^{c}\right]+
\left[\left[\hat{\Psi}^{c},\hat{\Psi}^{a}\right],\hat{\Psi}^{b}\right]+
\left[\left[\hat{\Psi}^{b},\hat{\Psi}^{c}\right],\hat{\Psi}^{a}\right]=0
\ee
and the commutator relation
\be
\left[\hat{\Psi}^{a},\hat{\Psi}^{b}\right]=-\left[\hat{\Psi}^{b},
  \hat{\Psi}^{a}\right]
\ee
which imply that
\be
U^{ab}_{e}d^{ec}+U^{ca}_{e}d^{eb}+U^{bc}_{e}d^{ea}=0,\label{cond1}
\ee
\be
d^{ab}=-d^{ba}\label{cond2}.
\ee
If we substitute the structure constants from
(\ref{cacon1})-(\ref{cacon3}) in (\ref{cond1}) we will find that the central
extensions of (\ref{gcomm}) can be written in terms of four constants $d_{i},
i=1,\ldots,4$
\be
d^{1,-1}_{m}&=&2(id_{4}+id_{3}m)\label{d}\\ d^{L,L}_{m}&=&d_{1}m^{3}+d_{2}m\\
d^{0,L}_{m}&=&d_{3}m^{2}+d_{4}m\label{d0l}\\ d^{0,0}_{m}&=&-imd_{3}.
\ee

 We assume now that $\hat{\cal Q}\equiv {\cal Q}_{P}$ is the ordered
version of ${\cal Q}$ in (\ref{Q2}) such that ${\cal Q}_{P}\left|
phys\right\rangle =0$. Notice that ${\cal Q}$ and ${\cal Q}_{P}$ are related
by the following equation
\bbe{oQ}
{\cal Q}={\cal Q}_{P}+A_{0}c^{0}_{0}+A_{L}c^{L}_{0}
\eee
where $A_{0}$ and $A_{L}$ are two ordering constants.

To check the nilpotency of ${\cal Q}_{P}$ we use the following trick
 \cite{MarnABRST,ISBE}.
First we define the operators $\tilde{\phi}^{i}_{n}$ by the equation
\bbe{tildephi}
\tilde{\phi}^{i}_{n}\equiv \{ b^{i}_{n},{\cal Q}_{P}\}
\eee
where $i=-1,0,1,L$. In the absence of anomalies the operators
$\tilde{\phi}^{i}_{n}$ act as gauge generators which preserve the
ghost number of the state on which they act
\cite{MarnABRST,MarnBRST}. Classically one can prove that the extended
constraints (\ref{tildephi}) satisfy the same algebra as the original
constraints (\ref{cacon1})-(\ref{cacon3}).

Using (\ref{Q2}) and (\ref{tildephi}) we find that the extended
constraints are given by the following relations
\be
\tilde{\phi}^{-1}_{m}&=&:\!\phi^{-1}_{m}\! :+ \sum_{-\infty}^{+\infty}:
[2ic^{1}_{k}b^{0}_{m-k}+ic^{0}_{k}b^{-1}_{m-k}-(m+k)c^{L}_{k}b^{-1}_{m-k}]:
\label{excon1}  \\
\tilde{\phi}^{0}_{m}&=&:\!\phi^{0}_{m}\! :+ \sum_{-\infty}^{+\infty}:
[ic^{1}_{k}b^{1}_{m-k}-ic^{-1}_{k}b^{-1}_{m-k}-mc^{L}_{k}b^{0}_{m-k}]:
-A_{0}\delta_{m}\label{excon2} \\
\tilde{\phi}^{1}_{m}&=&:\!\phi^{1}_{m}\! :- \sum_{-\infty}^{+\infty}:
[2ic^{-1}_{k}b^{0}_{m-k}+ic^{0}_{k}b^{1}_{m-k}+(m-k)c^{L}_{k}b^{1}_{m-k}]:\\
\tilde{\phi}^{L}_{m}&=&:\! \phi^{L}_{m}\! :- \sum_{-\infty}^{+\infty}:
[(k-m)c^{1}_{k}b^{1}_{m-k}+(k+m)c^{-1}_{k}b^{-1}_{m-k}+\nonumber \\
& &+kc^{0}_{k}b^{0}_{m-k}+(k+m)c^{L}_{k}b^{L}_{m-k}]:-A_{L}\delta_{m}
\label{excon4}
\ee
where  : : is the physical ordering  defined in section 2.3.

{}From the relations (\ref{com}) we can see that ordering ambiguities
arise only in the relations for $\tilde{\phi}^{0}_{0}$ and
$\tilde{\phi}^{L}_{0}$.  It is instructive to define the following ghost
operators
\be
\hat{G}^{i}_{0}&=&i: \sum_{k} c^{\hat{\imath}}_{k}b^{\hat{\imath}}_{-k}:
\label{gcon1}\\
\hat{G}^{i}_{L}&=&: \sum_{k} kc^{\hat{\imath}}_{k}b^{\hat{\imath}}_{-k}:
\ee
where $\hat{\imath}$ means that these indices are not summed over. Since
there are ordering ambiguities the action of these operators on the
vacuum will be given by the relations
\be
(\hat{G}^{i}_{0}-\beta^{i}_{0})\left| G\right\rangle =0,\\
(\hat{G}^{i}_{L}-\beta^{i}_{L})\left| G\right\rangle =0.\label{gcon2}
\ee
where $\beta^{i}_{0,L}$ are ordering constants. This is allowed since
the operators $\hat{G}^{i}_{0}$ and $\hat{G}^{i}_{L}$ commute with
each other for any $i$.  Notice that these
equations correspond to setting
$A_{0}=\alpha_{0}+\beta^{1}_{0}-\beta^{-1}_{0}$ and
$A_{L}=\alpha_{L}-\beta^{1}_{L}-\beta^{0}_{L}-\beta^{-1}_{L}-\beta^{L}_{L}$
in (\ref{excon2}) and (\ref{excon4}) respectively.

Since classically the extended constraints satisfy the same algebra as
the original constraints, the structure constants for the extended
constraint algebra are the same as the ones of the original algebra.
This, combined with the fact that the only operators for which we have
ordering ambiguities are again $\tilde{\phi}^{0}_{0}$ and
$\tilde{\phi}^{L}_{0}$ means, that we can use the results obtained previously
to write
\be
\left[  \tilde{\phi}^{1}_{m},\tilde{\phi}^{-1}_{n}\right] &=&2i
  \tilde{\phi}^{0}_{m+n}+2(i\tilde{d}_{4}+i\tilde{d}_{3}m)\delta_{m+n}
  \label{excona} \\
  \left[ \tilde{\phi}^{L}_{m},\tilde{\phi}^{L}_{n}\right] &=&(m-n)
    \tilde{\phi}^{L}_{m+n}+(\tilde{d}_{1}m^{3}+\tilde{d}_{2}m)\delta_{m+n}
    \label{d12} \\
    \left[ \tilde{\phi}^{0}_{m},\tilde{\phi}^{L}_{n}\right] &=&m
      \tilde{\phi}^{0}_{m+n}+(\tilde{d}_{3}m^{2}+\tilde{d}_{4}m)\delta_{m+n}
      \label{d34} \\
      \left[ \tilde{\phi}^{0}_{m},\tilde{\phi}^{0}_{n}\right] &=&
        -im\tilde{d}_{3}\delta_{m+n} \\
        \left[ \tilde{\phi}^{1}_{m},\tilde{\phi}^{0}_{n}\right]  &=&
          i\tilde{\phi}^{1}_{m+n}\\
          \left[ \tilde{\phi}^{-1}_{m},\tilde{\phi}^{0}_{n}\right]  &=&
            -i\tilde{\phi}^{-1}_{m+n}\\
            \left[ \tilde{\phi}^{1}_{m},\tilde{\phi}^{L}_{n}\right]   &=&
              (m+n)\tilde{\phi}^{1}_{m+n}\\
              \left[ \tilde{\phi}^{-1}_{m},\tilde{\phi}^{L}_{n}\right]   &=&
                (m-n)\tilde{\phi}^{-1}_{m+n}\label{excont}.
\ee

We combine all these equations in one, writing
\bbe{gextal}
[\tilde{\phi}^{i}_{m},\tilde{\phi}^{j}_{n}]= \sum_{s}
\sum_{k=-\infty}^{+\infty}U^{ij}_{s}(m,n,k)\tilde{\phi}^{s}_{k}+{\tilde
{d}}^{ij}_{m}\delta_{m+n}
\eee
where $i,j,s=-1,0,1,L$.

We can now calculate the BRST anomaly using a method described in
\cite{MarnABRST,ISBE}. There it is shown that

\be
{\cal Q}_{P}^{2}=\half \sum_{i,j}{\tilde {d}}^{ij}c^{i}_{m}c^{j}_{-m}.
\ee
So, if we substitute the values of the ${\tilde {d}}^{ij}$'s from
(\ref{excona})-(\ref{excont}) we will have that
\be
{\cal Q}_{P}^{2}&=&\tilde{d}_{1} \sum_{m}
\frac{m^{3}}{2}c^{L}_{m}c^{L}_{-m}\nonumber\\ &+&\tilde{d}_{2}
\sum_{m} \frac{m}{2}c^{L}_{m}c^{L}_{-m}\nonumber\\ &+&\tilde{d}_{3}
\sum_{m} (-i\frac{m}{2}c^{0}_{m}c^{0}_{-m}+m^{2}c^{0}_{m}c^{L}_{-m}+2i
mc^{1}_{m}c^{-1}_{-m}) \nonumber\\ &+&\tilde{d}_{4}
\sum_{m}(mc^{0}_{m}c^{L}_{-m}+2i c^{1}_{m}c^{-1}_{-m}).
\ee
The exact values of $\tilde{d}_{f}, f=1\ldots 4$ depend on the
vacuum and ordering we have used.

The simplest and safest method to determine these constants is to
calculate the matrix element of the commutators (\ref{excona})-(\ref{excont})
between the bra and ket vacuum.
{}From the relations (\ref{excon1})-(\ref{excon4}) we can see that
the ordered extended constraints satisfy the relations
\be
\tilde{\phi}^{-1}_{m}\left| 0\right\rangle &=&\left\langle 0\right|
\tilde{\phi}^{1}_{m}=0\hspace{.2in},\forall m\\
\tilde{\phi}^{0}_{m}\left| 0\right\rangle &=&\tilde{\phi}^{L}_{m}\left|
0\right\rangle =0\hspace{.2in},\forall m\neq 0.
\ee
As mentioned in  section  2.4 we  start from the hermitean ${\cal{Q}}_{H}$.
 Then by putting all terms in physical order we
can calculate the constants $\alpha_{0}$, $\alpha_{L}$, $\beta^{i}_{0}$ and
$\beta^{i}_{L}$ for this particular ordering. The calculation can be found in
the Appendix A.

The expectation value of the commutator (\ref{d12}) is
\bbe{fcon1}
\left\langle 0\right|[\tilde{\phi}^{L}_{m},\tilde{\phi}^{L}_{-m}]\left|
  0\right\rangle =2m\left\langle 0\right|\tilde{\phi}^{L}_{0}\left|
  0\right\rangle +\tilde{d}_{1}m^{3}+\tilde{d}_{2}m\nonumber\\
  \Rightarrow 0=2m(\alpha_{L}-\beta^{1}_{L}-\beta^{0}_{L}-\beta^{-1}_{L}-
  \beta^{L}_{L})-\tilde{d}_{1}m^{3}+\tilde{d}_{2}m\nonumber\\
  \Rightarrow \tilde{d}_{1}=0,\qquad
  \tilde{d}_{2}=-2(\alpha_{L}-\beta^{1}_{L}
  -\beta^{0}_{L}-\beta^{-1}_{L}-\beta^{L}_{L}).
\eee
In the same way from (\ref{d34}) we have
\bbe{fcon2}
\left\langle 0\right|[\tilde{\phi}^{0}_{m},\tilde{\phi}^{L}_{-m}]\left|
  0\right\rangle =m\left\langle 0\right|\tilde{\phi}^{0}_{0}\left|
  0\right\rangle +\tilde{d}_{3}m^{2}+\tilde{d}_{4}m\nonumber\\
  \Rightarrow 0=m(\alpha_{0}+\beta^{1}_{0}-\beta^{-1}_{0})+\tilde{d}_{3}m^{2}
  +\tilde{d}_{4}m=0\nonumber\\
  \Rightarrow \tilde{d}_{3}=0,\qquad
  \tilde{d}_{4}=\beta^{-1}_{0}-\beta^{1}_{0}
  -\alpha_{0}.
\eee
So using the results from the
appendix A, in (\ref{fcon1}) and (\ref{fcon2}), we have that
\be
\tilde{d}_{1}=0,&\qquad &\tilde{d}_{2}=(D+4)\lim_{N\to +\infty}\left(
\sum_{k=-N}^{+N}k \right)\label{fconst1}\\  \tilde{d}_{3}=0,&\qquad &
\tilde{d}_{4}=\frac{i}{4}(D-2)\lim_{N\to +\infty}\left( \sum_{k=-N}^{+N}1
\right)\label{fconst2}.
\ee

On the other hand when we calculate the expectation value of the commutator
(\ref{excona})  we  find the following
\bbe{fcon3}
\tilde{d}_{m}^{1,-1}+2i\left\langle 0\right|\tilde{\phi}^{0}_{0}\left|
0\right\rangle &=&\left\langle 0\right|[\tilde{\phi}^{1}_{m},
\tilde{\phi}^{-1}_{-m}]\left| 0\right\rangle =-\left\langle 0\right|
\tilde{\phi}^{-1}_{0}\tilde{\phi}^{1}_{0}\left| 0\right
\rangle \nonumber\\
&=&-\left\langle 0\right|\phi^{-1}_{-m}\phi^{1}_{m}\left|
0\right\rangle -2 \sum_{k,l}\left\langle
0\right|c^{1}_{-k-m}b^{0}_{k}c^{0}_{-l+m}b^{1}_{l}\left|
0\right\rangle \nonumber\\ &=&\frac{1}{2}(D+2)\lim_{N\to
+\infty}\left( \sum_{k=-N}^{+N}1\right ) -2\lim_{N\to +\infty}\left(
\sum_{k=-N}^{+N}1\right )\Rightarrow\nonumber\\
\tilde{d}_{m}^{1,-1}&=& 0.
\eee
But $\tilde{d}_{m}^{1,-1}, \tilde{d}_{3}$ and $\tilde{d}_{4}$ have to satisfy
the  requirement (\ref{d}) which comes from the Jacobi identities. This means
that the following relation should be satisfied
\bbe{Jacobi}
0=-\frac{1}{2}(D-2)\lim_{N\to +\infty}\left( \sum_{k=-N}^{+N}1 \right).
\eee
 This holds {\it only when} $D=2$.

Substituting the results (\ref{fconst1}),(\ref{fconst2}) and
(\ref{Jacobi}) in (\ref{Q2}) we obtain that
\be
{\cal Q}_{P}^{2}=\half (2+4)\lim_{N\to +\infty}\left(
\sum_{k=-N}^{+N}k \right) \sum_{m} mc^{L}_{m}c^{L}_{-m}.
\ee

Finally using the  world-sheet parity invariant regularization scheme which
is discussed in the appendix we take $(\sum k\to 0)$ and get
\bbe{anoma}
{\cal Q}^{2}_{P}=0
\eee
which, as shown before, holds only in the critical dimension $D=2$.

We thus recover a consistent theory in the extended phase space of ghost and
matter fields, in two space time dimensions. So we see that there are
obstructions to preserve the conformal symmetry of the classical tensionless
string at the quantum level in any other dimension, in agreement with the
result in
\cite{BigT}.

It is interesting to observe that the ghosts are vital for the theory to be
independent of ordering prescription. This is illustrated in Appendix B using
the usual tensile bosonic string as an exemple.

It is also possible to understand why the Jacobi identities do not
hold using the local field language of the subsections 2.1-2.4.
In general, to construct a quantized algebra one has to regularize ordering
ambiguities. However, for a proper quantized algebra to exist, commutators
of generators should be identified as linear combinations of the generators
{\em in the limit when $\epsilon\rightarrow 0$}. Schematically this could be
written
  \be
  \left[\phi_{\epsilon}^{a},\phi_{\epsilon}^{b}\right]=U^{ab}_{c}
    \phi^{c}_{\epsilon}+{\cal O}(\epsilon).
  \ee
The generators $\phi_{\epsilon}$ are $\epsilon$ dependent but have a well
defined, non-singular, $\epsilon$ independent limit when
$\epsilon\rightarrow 0$. In particular this means that they must be
physically ordered. The terms which are not identifiable as generators of
the algebra are all proportional to positive powers of $\epsilon$ and thus
vanish when $\epsilon\rightarrow 0$. If the $\epsilon$ dependence cannot be
removed, there is an obstruction to representing the quantized algebra as
well-defined linear operators. Algebraically such problems show up when
commuting ``once more'', \ie in triple commutators. The test that triple
commutators are algebraically consistent is the Jacobi identities. Thus,
regularization problems of the present kind cause the Jacobi identities to
fail.

It is also possible to understand technically how this comes about. The terms
that one throws away in the present construction, \ie terms that are
proportional to powers of $\epsilon$, may contribute in commutation
relations. To appreciate this, recall what was said in subsection 2.3. In
formula (\ref{PSplit}) we saw that changing the order of $X$ and $P$ fields
brings out nontrivial negative powers of $\epsilon$, the simplest example
being
\be
 X^{M}(\sigma)P^{N}(\sigma)=P^{N}(\sigma)X^{M}(\sigma) + \frac{i}{\epsilon}
 \eta^{MN}\delta_{1}(0).
\ee
With this in mind, one may check that there exist commutators between
physically ordered terms, seemingly proportional to positive powers of
$\epsilon$, which give contributions in further commutators that do not vanish
when $\epsilon$ goes to zero. As an illustration, consider a commutator
between two typical terms arising in the calculation of the algebra
\be
\left[ \epsilon^{2}X(\sigma)\cdot X^{\prime\prime}(\sigma),P^{2}
  (\sigma^{\prime})\right] =
  \frac{2(D+2)}{\epsilon}\delta^{\prime\prime}_{1}
  (0)\delta_{\epsilon}(\sigma-\sigma^{\prime})+{\cal O}(\epsilon).
\ee
This demonstrates that throwing away ${\cal O}(\epsilon)$ terms {\em before}
calculating the Jacobi identities means that we in general risk throwing away
terms that contribute in the $\epsilon\rightarrow 0$ limit. The
$\epsilon\rightarrow 0$ gauge algebra is only consistently represented when
the Jacobi identities are satisfied, \ie for $D=2$.

\begin{flushleft}
\section{Discussion}
\end{flushleft}

In this section we collect some thoughts on the result presented
above.

The discovery of $D=2$ as a critical dimension in the quantum theory
prompts us to take a
closer look at that case also in the classical theory.
We will show below that
in this dimension the solutions to the equations of motion are simply
massless particles.

The critical dimension $D=2$ is furthermore compared with the result of
\art{BigT} where no critical dimension was discovered.

We also list some topics for future considerations.\\

\subsection{Two dimensions, a special case}

We start from the action (\ref{LDAC}) and derive
the $V^{\alpha}$ equations
\bbe{TDEQ}
\underline{\dot X}\cdot\partial\underline{X}=0\quad
\underline{X'}\cdot\partial\underline{X}=0.
\eee
Here
spacetime vectors are underlined and
\bbe{last}
\partial \equiv
V^\alpha\partial_\alpha.
\eee
{}From (\ref{TDEQ}) it follows that
$(\partial\underline{X})^2=0$, which, in $D=2$ implies that
{\footnote{In this context, we disregard the solutions
    $\partial\underline{X}=0$ which carry zero momentum.}}
$\partial\underline{X}=\alpha (\xi)\underline{e}$ with $\underline{e}$ a
vector in one of the two null directions.  Substituting this
information back into (\ref{TDEQ}) we conclude that also
$\underline{\dot X}$ and $\underline{X'}$ are proportional to
$\underline{e}$. Denoting the proportionality functions by $\beta
(\xi)$ and $\gamma (\xi)$ respectively, we find two expressions for
$\underline{X}$:
\bbe{Xrel}
\underline{X}=\underline{e}\int{\beta (\xi) d\tau}
+\underline{f}(\sigma ) \cr
\underline{X}=\underline{e}\int{\gamma (\xi) d\sigma}
+\underline{g}(\tau ).
\eee
Taking the $\sigma$-derivative of the first expression and
comparing it to $\underline{X'}=\gamma (\xi)\underline{e}$ we
determine $\underline{f}$. Thus we find:
\bbe{Xpart}
\underline{X}=\underline{e}\int{\gamma (\xi) d\sigma}
+\underline c\equiv \Gamma (\xi)\underline{e}+\underline c,
\eee
where $\underline c$ is a constant vector. We may now make a $2D$
world-sheet coordinate transformation $\Gamma (\xi)
\to \tau$ to rewrite (\ref{Xpart}) as the expression for a
massless particle:
\bbe{tau}
\underline{X}=\tau\underline{e}+\underline c.
\eee

Note that, perhaps somewhat surprisingly, we arrived at (\ref{tau})
using the $V^{\alpha}$ equations only. The $\underline{X}$ equation,

\bbe{Xation}
\partial_\alpha (V^\alpha \partial\underline{X}) =0,
\eee
will determine $V^{\alpha}$ instead. Using (\ref{Xpart}), the relation
(\ref{Xation}) becomes
\bbe{dve}
\partial_\alpha(V^\alpha\partial\Gamma )\underline{e}=\underline{0},
\eee
with (local) solution
\bbe{Vsln}
V^\alpha\partial\Gamma &=\epsilon ^{\alpha \beta}\partial _\beta h\cr
\Leftrightarrow
\epsilon _{\alpha \beta}V^\beta \partial\Gamma &=\partial _\alpha
h.
\eee
Here $h(\xi )$ is a scalar density which satisfies $\partial h=0$.\\

Clearly the relation \ekv{tau} is insensitive to an arbitrary
$\sigma$-coordinate transformation $\sigma \to \tilde\sigma
(\xi )$. This is the manner in which the residual (Virasoro)
symmetry arises in two dimensions.\\

In the gauge used in (\ref{tau}) the solution (\ref{dve}) takes on the
form
\bbe{taug}
(V^0)^2=h', \qquad V^1V^0=-\dot{h}.
\eee
As long as $h'$ is non-zero this is equivalent to
\bbe{equv}
V^0=\sqrt{h'}, \qquad V^1=-\dot{h}/\sqrt{h'}.
\eee
We see that the
$V^\alpha$'s are determined by one field $h$. From the
first relation in (\ref{taug}) it is clear that $h' \ge 0$. For a
closed string and a globally defined $h$, we must have periodicity
$h(\sigma + 2\pi)=h(\sigma)$, and consequently $h' \ge 0 \Rightarrow h'= 0$.
{}From (\ref{taug}) we see that this leaves $V^1$ undetermined instead. In each
case the indeterminacy reflects the residual gauge symmetry and does not
represent physical degrees of freedom. For $V^\alpha$ of trivial
topology conventional Hamiltonian treatment of the system described
above gives the degrees of freedom of a massless particle.

\subsection{Comparision to previous results}

Having thus displayed the classical $2D$ structure of the theory we
see how the discovery of a critical dimension is in agreement with
previous results. Namely, in \art{BigT} the physical degrees of
freedom that are quantized are the transversal ones. This excludes
$D=2$ from the outset and explains why no critical dimension is found
there. Furthermore, in \art{BigT} it is shown that the conformal
invariance of the massless {\em particle} survives quantization. The
proof is again for $D>2$, but it is reasonable to expect this result
to carry over to $D=2$.

It is also proper to compare our calculation to other known results to see if
our methods are consistent. The first thing one thinks of is the
Lorentz symmetry of the tensionless string. It is well known that,
ignoring the space time conformal invariance, quantization goes
through with no problems at all, in any space time dimension
\cite{liraspsr,gararual}. To compare our calculation with that result,
we notice that the structure of two of our constraints are exactly
similar to the ordinary constraints in the tensionless theory.
Therefore we will immediately be able to compare results if we,
everywhere in our calculation, just disregard everything that
has to do with our two extra constraints, $\phi^{0},\phi^{1}$ and our two
extra dimensions. We may of course keep our choice of vacuum, and accordingly
our physical ordering, since the discussion of these matters is generally
valid for tensionless strings. We find a BRST charge
\bbe{lBRST}
  \tilde{\cal {Q}}_{P}=\int d\sigma \left( \phi^{-1}c^{-1}+\phi^{L}c^{L}
+i\partial c^{-1}c^{L}b^{-1}+ib^{L}\partial c^{L}c^{L}
+\frac{ia}{2\epsilon}\partial c^{L}+{\cal O}(\epsilon)\right),\nonumber\\
\mbox{}
\eee
and we may check that $\tilde{\cal Q}_{P}^{2}$ is proportional to an integral
of a total derivative and thus vanishes. Furthermore, there are no problems
with the Jacobi identities, neither for bosonic nor extended constraints.

The other model which bears resemblance to our case and with which we
would like to compare is the conformal particle \cite{MarnBRST}, being
essentially a model containing only the zero modes of the $\phi$
constraints. In this case, the quantized theory is consistent
regardless of the dimensionality of the ambient space. We may thus
conclude that it is the richness of the state space, a consequence of the
extendedness of strings which causes the problems; in two dimensions the
extendedness effectively vanishes, and in higher dimensions we have a
consistent theory if we only look at the zero modes of the constraints. This
conclusion also fits with the results of \cite{BigT}. We thus find that our
method gives results that agree with those arrived at by other routes.\\

In this article we have not used the philosophy of \cite{BigT}, imposing
restrictions on the physical state space to avoid the problems in $D\neq 2$.
It would be interesting to see where this would lead and if we could recover
the results that physical states should be space time diffeomorphism
invariant. The natural generalization in our case would be to require that
${\cal Q}^2=0$ only on physical states. Since commutators between operators
annihilating physical states should also annihilate physical states, one would
by following this route derive a large number of constraints on physical
states, possibly leading to the result of \cite{BigT}. There is however
another more difficult problem one would have to tackle in adopting this
philosophy, namely the failure of the Jacobi identities to close in $D\neq 2$.
This is a problem one would have to solve to be able to carry out the program
pursued in \cite{BigT}.

\subsection{Outlook}

In this paper we have studied BRST-quantization of the conformal string and
found  obstructions to quantization except in two space-time dimensions.
A novel feature of our treatment is the important role played by the Jacobi
identities. We have also emphasized the necessity of chosing a correct vacuum
and we have explained how to reconcile our results with previous ones in the
literature. It remains to analyze the structure of the operator anomalies
and see if it is again possible to view them as giving restrictions on the
physical states of the theory. If that {\em is} possible,we must also answer
the question of how this leads to the "topological state space" result of
\art{BigT}.

Another obvious avenue of investigation is to  turn to the $T\to 0$ limit of
the superstring and the spinning string. In particular, applying the
techniques of the present paper to the superstring presents an interesting
challenge.

It should furthermore be interesting to take a closer look at the
two-dimensional case and see if it is possible to explicitly construct the
quantum theory there.

\bigskip
\begin{flushleft}
{\bf Acknowledgments:} We thank Johan Grundberg, Stephen Hwang, Martin Ro\v
cek and Rodanthy Tzani for useful comments and fruitful discussions. The
research of UL was supported in part by NFR under contract No F-AA/FU
04038-312  and by NorfA under contract No 94.35.049-O.
\bigskip
\end{flushleft}

\eject

\renewcommand{\thesection}{Appendix \Alph{section}}
\setcounter{section}{0}
\renewcommand{\theequation}{\Alph{section}.\arabic{equation}}
\setcounter{equation}{0}
\section{Ordering constants}
Here we
calculate the constants
$\alpha_{0},\alpha_{L},\beta^{i}_{0}$ and $\beta^{i}_{L}$. To do this we use
the following regularization scheme
\bbe{reg}
\sum^{+\infty}_{-\infty} \rightarrow \lim_{N \to +\infty}\sum^{+N}_{-N}.
\eee
This means that instead of calculating the infinite sums we first calculate
the sums using finite limits $+N,-N$ and at the end we take the $N \to
+\infty$ limit. In this way we see that
\be
\left( \sum_{k=-\infty}^{+\infty}k\right) \to 0
\ee
while on the other hand $\left( \sum_{k=-\infty}^{+\infty}1\right)$ remains
divergent.

Starting from the relations (\ref{ccon2}),(\ref{ccon4}) and
(\ref{ord1}),(\ref{ord2}) for the physical  part of the  extended constraints
and (\ref{gcon1})-(\ref{gcon2}) for the ghost part we use the ordering we
defined  in section 2.3 to find the following
\be
\hat{\phi}^{0}_{0}\left| 0\right\rangle &=&\frac{1}{4} \sum(x_{-k}\cdot
p_{k}+p_{k}\cdot x_{-k})\left| 0\right\rangle\nonumber\\
&=&\half \sum (x_{-k}\cdot
p_{k})
\left| 0\right\rangle -\frac{i}{4}(d+2)\sum 1 \left| 0\right\rangle \nonumber\\
& &\Rightarrow \alpha_{0}=-\frac{i}{4}(d+2)\left(
\sum_{k=-N}^{+N} 1\right)\label{app1}\\
\hat{\phi}^{L}_{0}\left| 0\right\rangle &=&\frac{i}{2} \sum k(x_{-k}\cdot
p_{k}-p_{k}\cdot x_{-k})\left| 0\right\rangle \nonumber\\
&=&i\sum kx_{-k}\cdot p_{k}\left|
0\right\rangle -\half (d+2) \sum k \left| 0\right\rangle \nonumber\\
& &\Rightarrow \alpha_{L}=-\half (d+2) \left( \sum_{k=-N}^{+N}k\right)
\label{app2}\\
\hat{G}^{1}_{0}\left| 0\right\rangle &=&\frac{-i}{2}\sum (b^{1}_{-k}c^{1}_{k}
-c^{1}_{k}b^{1}_{-k})\left| 0\right\rangle = -i \sum
b^{1}_{-k}c^{1}_{k}\left|
0\right\rangle +\frac{i}{2} \sum 1\left| 0\right\rangle \nonumber\\
& &\Rightarrow \beta^{1}_{0}=\frac{i}{2} \left(
\sum_{k=-N}^{+N} 1\right) \label{app3}\\
\hat{G}^{i\neq 1}_{0}\left| 0\right\rangle &=&\frac{i}{2} \sum
(c^{\hat{i}}_{k}b^{\hat{i}}_{-k}-b^{\hat{i}}_{-k}c^{\hat{i}}_{k})\left|
0\right\rangle =i \sum c^{\hat{i}}_{k}b^{\hat{i}}_{-k}\left| 0\right\rangle
-\frac{i}{2} \sum 1 \left| 0\right\rangle \nonumber\\
& &\Rightarrow \beta^{i\neq 1}_{0}=-\frac{i}{2}\left(
\sum_{k=-N}^{+N} 1\right) \label{app4}\\
\hat{G}^{1}_{L}\left| 0\right\rangle &=& -\half \sum k(b^{1}_{-k}c^{1}_{k}+
c^{1}_{k}b^{1}_{-k})\left| 0\right\rangle = -\sum kb^{1}_{-k}c^{1}_{k}\left|
0\right\rangle -\half \sum k\left| 0\right\rangle \nonumber\\
& &\Rightarrow \beta^{1}_{L}=-\half \left( \sum_{k=-N}^{+N}k\right)
\label{app5}\\
\hat{G}^{i\neq 1}_{L}\left| 0\right\rangle &=& \half \sum k(c^{\hat{i}}_{k}
b^{\hat{i}}_{-k}+b^{\hat{i}}_{-k}c^{\hat{i}}_{k})\left| 0\right\rangle =
\sum k(c^{\hat{i}}_{k}b^{\hat{i}}_{-k})\left| 0\right\rangle +\half \sum
k\left| 0\right\rangle \nonumber\\
& &\Rightarrow \beta^{i\neq 1}_{L}=\half \left(
\sum_{k=-N}^{+N}k\right).\label{app6}
\ee

We should comment here that all these constants are divergent in contrast
to the usual tensile case where the ordering constants have finite values.

\setcounter{equation}{0}
\section{Some remarks on the critical dimension}

 Let us
calculate the Virasoro algebra for the usual tensile
string using the cutoff regularization (\ref{reg}).

  We have that
\bbe{comm}
\left[ \phi^{L}_{m},\phi^{L}_{-m}\right]=\lim_{N \to +\infty}\left\{
\begin{array}{cccc}
m\alpha^{2}_{0}+2m\sum^{N -m}_{n=1}
\alpha_{-n}\cdot\alpha_{n}+\frac{D}{12}m(m^{2}-1)\\ +\sum^{N}_{k=N
-m+1}(m-k)\alpha_{-k}\cdot\alpha_{k}\\
\\
m\alpha^{2}_{0}+2m\sum^{N -m}_{n=1}
\alpha_{n}\cdot\alpha_{-n}-\frac{D}{12}m(m^{2}-1)\\ +\sum^{N}_{k=N
-m+1}(m-k)\alpha_{k}\cdot\alpha_{-k}
\end{array}\right.
\eee
where we have expressed the result in two different orderings. In the first
line all the positive modes are put to the right and all negative modes to
the left. The second line is ordered in the opposite way. Reordering the
first line will reproduce the second.

We can take  the limit $N \to +\infty$ by throwing away the last terms in
eqs. (\ref{comm}). Their matrix element between states of fixed mass vanish
as $N \to +\infty$. The commutator then becomes
\be
\left[ \phi^{L}_{m},\phi^{L}_{-m}\right]=\left\{
\begin{array}{cc}
m\alpha^{2}_{0}+2m\sum^{+\infty}_{n=1}
\alpha_{-n}\cdot\alpha_{n}+\frac{D}{12}m(m^{2}-1)\\
\\
m\alpha^{2}_{0}+2m\sum^{+\infty}_{n=1}
\alpha_{n}\cdot\alpha_{-n}-\frac{D}{12}m(m^{2}-1)
\end{array}\right.
\ee
If we
now reorder the first term we will not reproduce the second. To have
relations independent of the ordering prescription for $D\neq 0$ we have to
introduce ghosts. Then the Virasoro operators become the extended Virasoro
operators
\be
\tilde{\phi}^{L}_{m}=\left\{Q,b_{m}\right\}=\phi^{L}_{m}+\phi^{Lc}_{m}=
\phi^{L}_{m}+\sum^{+\infty}_{n= -\infty} (m-n)b_{m+n}c_{-n}
\ee
Using the relation (\ref{reg}) the algebra in the extended space
becomes
\be
\left[ \phi^{L}_{m},\phi^{L}_{-m}\right]=\lim_{N \to +\infty}\left\{
\begin{array}{lccclccc}
m\alpha^{2}_{0}+2m\sum^{N -m}_{n=1}
\alpha_{-n}\cdot\alpha_{n}+\frac{D}{12}m(m^{2}-1)\\
+2m\sum^{N
-m}_{k=1}k(b_{-k}c_{k}+c_{-k}b_{k})+\frac{1}{6}(m-13m^{3})\\ +\sum^{N}_{k=N
-m+1}(m-k)\alpha_{-k}\cdot\alpha_{k}\\ +\sum_{k=N
-m+1}^{N}(2m-k)(m+k)(c_{-k}b_{k}+b_{-k}c_{k})\\
\\
m\alpha^{2}_{0}+2m\sum^{N -m}_{n=1}
\alpha_{n}\cdot\alpha_{-n})-\frac{D}{12}m(m^{2}-1)\\
+2m\sum^{N
-m}_{k=1}k(b_{k}c_{-k}+c_{k}b_{-k}-\frac{1}{6}(m-13m^{3})\\ +\sum^{N}_{k=N
-m+1}(m-k)\alpha_{k}\cdot\alpha_{-k}\\ +\sum_{k=N
-m+1}^{N}(2m-k)(m+k)(c_{k}b_{-k}+b_{k}c_{-k})\\
\end{array}\right.
\ee

In the limit $N\rightarrow +\infty $ this relation becomes
\bbe{fcomm}
\left[ \phi^{L}_{m},\phi^{L}_{-m}\right]=\left\{
\begin{array}{cccc}
m\alpha^{2}_{0}+2m\sum^{+\infty}_{n=1}
\alpha_{-n}\cdot\alpha_{n}+2m\sum^{+\infty}_{k=1}k(b_{-k}c_{k}+c_{-k}b_{k})\\
+\frac{D}{12}m(m^{2}-1)+\frac{1}{6}(m-13m^{3})\\
\\
m\alpha^{2}_{0}+2m\sum^{+\infty}_{n=1}
\alpha_{n}\cdot\alpha_{-n}+2m\sum^{+\infty}_{k=1}k(b_{k}c_{-k}+c_{k}b_{-k})\\
-\frac{D}{12}m(m^{2}-1)-\frac{1}{6}(m-13m^{3})\\
\end{array}\right.
\eee

Using $\zeta$-function regularization
$\sum^{+\infty}_{n=1}n=-\frac{1}{12} $ we can show that the two branches of
(\ref{fcomm}) are equal only when
\be
\frac{m}{6}(D-2)+\frac{1}{6}D(m^{3}&-&m)+\frac{1}{3}(m-13m^{3})=0\nonumber\\
\Rightarrow D&=&26
\ee
So independence of ordering prescription requires that $D=26$.

We can study the conformal string in the same manner. In this case we will
have
\be
\left[\phi^{0}_{m},\phi^{L}_{-m}\right]=\lim_{N \to +\infty}\left\{
\begin{array}{l}
\frac{1}{2}m\sum^{N -m}_{-N +m} x_{k}\cdot p_{-k}\\
+\frac{1}{2}\sum^{N}_{N -m+1}(m-k)x_{k}\cdot p_{-k}\\
+\frac{1}{2}\sum
^{-N +m-1}_{-N}kx_{k}\cdot p_{-k}\\
\\
\frac{1}{2}m\sum^{N -m}_{-N +m} p_{-k}\cdot x_{k}\\
+\frac{1}{2}\sum^{N}_{N -m+1}(m-k)p_{-k}\cdot x_{k}\\
+\frac{1}{2}\sum
^{-N +m-1}_{-N}kp_{-k}\cdot x_{k}
\end{array}
\right.
\ee

So in the $N \to +\infty$ limit the two branches are different. Using
the previous procedure we extend the space to include ghosts and we
find ordering independent  results only when $D=2$.

\eject

 \end{document}